# Gaseous Radiochemical Method for Registration of Ionizing Radiation and Its Possible Applications in Science and Industry

*S.G. Lebedev*, *V.E. Yants*

Institute for Nuclear Research of Russian Academy of Sciences
60th October Anniversary Prospect, 7a
117312, Moscow, Russia

## Abstract

This work presents a new possibility of registration of ionizing radiation by the flowing gaseous radiochemical method (FGRM). The specified method uses the property of some solid crystalline lattice materials for a free emission of radioactive isotopes of inert gas atoms formed as a result of nuclear reactions. Generated in an ampoule of the detector, the radioactive inert gases are transported by a gas-carrier into the proportional gas counter of the flowing type, where the decay rate of the radioactive gas species is measured. This quantity is unequivocally related to the flux of particles (neutrons, protons, light and heavy ions) at the location of the ampoule. The method was used to monitor the neutron flux of the pulsed neutron target "RADEX" driven by the linear proton accelerator of INR RAS. Further progress of the FGRM may give rise to possible applications in nuclear physics, astrophysics and medicine, in the nondestructive control of fissionable materials, diagnostics of thermonuclear plasma, monitoring of fluxes and measurement of spectra of bombarding particles.

*Corresponding author Tel.: +7-95-3340714
*E-mail address:* lebedev@inr.ru

## 1. Introduction

For the on-line registration (monitoring) of ionizing radiation a flowing gaseous radiochemical method (FGRM) was proposed earlier [1-5]. The method was applied to register decays with the participation of neutrinos in a chlorine-argon detector [6] and also for the measurement of the background of fast neutrons with fluxes of the order of $\sim 10^{-7}$ $n/cm^2/day$ in a gallium-germanium detector [7]. The procedure is based on the nuclear reactions of neutrinos with chlorine and gallium, producing radioactive gases. A feature of the mentioned measurements was the exposition of large quantities (*100-1000 kg*) of target material in the fluxes of neutrinos and neutrons with subsequent condensation of the inert gas at low temperatures. For the determination of the quantity of radioactive gas atoms produced, the condensate was placed in a stationary proportional counter where all radioactive gaseous atoms were counted.

The FGRM described here differs from the method outlined above in the sense of a property of the crystalline lattice of some solid substances for a free release of radioactive inert gas atoms formed as the result of nuclear reactions. After the formation of the radioactive inert gas isotope in an ampoule loaded with target material, the released gas is transported by a gas-carrier into the proportional gas counter of a flowing type. Here the decay speed of the radioactive inert gas nuclei is measured; it is unequivocally proportional to the flux of ionizing radiation in the place of the target material in the ampoule. A feature of the offered method is the registration of fluxes of ionizing radiation with the help of the proportional flowing counter that allows executing the measurements in a real time scale.

In a stationary mode of irradiation of an ampoule (at a constant flux of ionizing radiation $\Phi$ and at a constant consumption $L$ of carrier-gas) all newly formed activity over equilibrium leaves the volume of an ampoule by a current of gas-carrier. Therefore, the specific activity of the gas-carrier is connected to the flux of ionizing radiation by a ratio:

$$I = \frac{V_C P_C N_0 \sigma \Phi}{1.44 t L}, \qquad (1)$$

where $I$ is the quantity of decays *per second* of radioactive inert gas in the counter, $V_C$ the "alive" volume of the counter in $cm^3$, $P_C$ the pressure in the counter in physical atmospheres, $N_0$ the number of atoms of target material in an ampoule, $t$ the half-life of a radioactive inert gas in *sec*, $\sigma$ the cross-section of a nuclear reaction for the formation of a radioactive isotope of an inert gas averaging on a spectrum of ionizing radiation, $\Phi$ the flux of ionizing radiation in $cm^{-2}sec^{-1}$ and $L$ the consumption of transport gas in $cm^3/sec$. Eq. (1) is applicable at $V_a/L \ll$ t, where $V_a$ is the volume of an ampoule with target material.

As can be seen from Eq. (1), the described method of registration of ionizing radiation is absolute and does not require calibration, as all quantities included in Eq. (1) can be measured with high accuracy. Apart from this it is possible to enumerate the other advantages of the specified method:

- Tolerance to gamma-quanta (the counters of decays of radioactive inert gases can be removed far enough and can be shielded);
- The short response time on the change of a flux of ionizing radiation (it is determined by the speed of gas exchange in an ampoule and the respective length and diameter of a transport pipeline);
- Absence of moving parts, and, as a consequence, the simplicity and reliability in operation. (Unlike to other circulating schemes, in which the active substance follows a contour, the full activity here is equal to an integral along a contour.) The degree of formation of radioactive inert gases is only determined by the fluence of ionizing radiation in the ampoule position.

- In the case of a solid state of the target material the vapor pressure is small, therefore no target material is transported away by a carrier-gas;
- The absence of a liquid phase automatically excludes evaporation or splashing of target material depending on the carrier-gas flow and facilitates a free regulation of the gas exchange speed.

The given method was applied to monitor a neutron flux of the pulsed neutron target "RADEX" [8-10] with illumination from the linear proton accelerator of the Moscow Meson Facility of INR RAS. The sketch of the installation "RADEX" is shown in a cross sectional view in Fig. 1 in ref. [9]. The proton beam passes through an aluminum first wall and falls in an active zone of the neutron target, assembled from tungsten plates with titanium coatings, cooled by light water. Inside an active zone on a depth of *~4 m* from a surface and at a distance of *~40 mm* from the first wall, the cylindrical irradiation channel with a diameter of *73 mm* is located . The irradiation channel is completely autonomous. It can be evacuated or equipped with an own cryogenic system [11], cooled by water, liquid metal or gas. This channel is used for the radiation tests of the standard samples of perspective alloys irradiated both with purely neutron spectra and in mixed spectra of protons and neutrons [8-10].

Fig. 2 represents a spectrum of decays in the proportional counter for the experiments at the RADEX facility. The *k*-peak (appropriate to the energy of *2,9 keV*) of $^{37}Ar$ decays is precisely seen in the vicinity of the 280 analyzer's channel. The low energy peak is stipulated by a beta-decay of $^{41}Ar$. From Fig. 2 it is visible that the spectrum from a beta-decay falls down smoothly with increasing energy, therefore, the contributions from $^{37}Ar$ and $^{41}Ar$ can be resolved effectively, demonstrating an example of simple spectrometry of fast neutrons at the installation RADEX.

During the time following the first experiments, the research activity on the probable applications of the specified method in the various areas of fundamental and applied sciences was continued [12]. Some findings are described below.

## 2. Precise neutron flux monitoring

The results obtained in the experiments at the RADEX facility and presented in Fig. 2 demonstrate the high sensitivity and selectivity of the FGRM for the monitoring of neutrons.

The case history of a neutron flux monitoring at the RADEX neutron target with the help of a FGRM detector is adduced in Fig. 3. Here one can see the relation of the decay counts vs. time during three days of continuous registration in an automatic mode. The dark colored areas correspond to the presence of a proton beam on the target. In the RADEX facility the linear proton accelerator drives the neutron target. Therefore all instabilities or jumps of the proton beam give rise to corresponding instabilities in the neutron flux. As can be seen from Fig. 3 the FGRM detector reproduces all changes of the neutron flux in the RADEX irradiation channel. The neutron pulse vertical edges are not delayed. This kind of behavior gives an idea of using the FRGM detector for precise monitoring of the time behavior of the neutron flux, for example, in the experiments on absolute neutron flux determinations [13] within the frame of the neutron lifetime project.

Of course, the problem of absolute neutron flux determination is not so simple as to be solved by the help of FGRM alone. First of all, the accuracy of data on the cross section of neutron interactions with nuclei are on the average not so high *(~10-20%)* to provide the required precision in the neutron lifetime experiment *(0.1%)*. So there is strong evidence needed in the cross calibration of measurements of various types of detectors. However, it is believed that the FGRM may help to increase the precision control of the time behavior of a neutron flux. Some of the arguments in support of this statement are listed below.

To illustrate the possibility of the FGRM for precise neutron monitoring let us consider the processes happening in the course of neutron flux measurements. A scheme of the real-time neutron-flux monitoring performed with the RADEX facility based upon the above approach is presented in Fig. 4. Powdered target material of $CaC_2O_4$ (calcium-oxalate) of *32 g* mass is enclosed in a steel ampoule placed in the neutron flux at the bottom of the irradiation channel. The carrier gas *(He)* from a vessel flows in *3-mm*-diameter stainless-steel tubes through a pressure regulator, a flow regulator and a flow-meter. It then travels through the ampoule, picking up the $^{37}Ar$ produced in the reaction $^{40}Ca(n,\alpha)^{37}Ar$ and upon passing through the cocks and flow-meters it enters a proportional counter located on top of the biological shielding. Gas flow in the pipeline may be subjected to stochastic fluctuations. Therefore the gas consumption must be a constant with high precision. This can be realized by means of a series connection of flow-meters. To reach the precision of *0.1%* about ten flow-meters with a measurement precision of *1%* should be used.

The counter is a cylindrical quartz bulb with a thin graphite layer deposited on its inner wall. This layer is the cathode of the ionization chamber. The anode is a thin tungsten wire running along the cylinder axis. The counter operates as follows: The carrier gas delivers the radioactive gas $^{37}Ar$ to the counter. Argon atoms decay inside the counter, emitting Auger electrons. The range of Auger electrons in the carrier gas is a few microns; therefore, the volume of the wall layer accounts for *~0.001%* of the entire counter volume of $\sim 60\ cm^3$. Under these conditions the counter detects decay processes with approximately 100% efficiency. As the $^{37}Ar$ is an inert (noble) gas, the absorption and pressure losses in the pipeline are negligible.

The counter is placed inside a box formed by sectional blocks of a *2-m*-thick concrete shielding. In addition, the counter is enclosed in a local lead shielding to suppress the background from scattered neutrons and gamma rays. At the exit from the irradiation channel, the stainless-steel tubes are joined to polyethylene tubes with a *4-mm* i.d. for the ease of hookup. The total length of the tubing is *~10 m*, and the transport time for the inert gas to travel from the ampoule to the counter (i.e. the signal-delay time) at a $1\text{-}cm^3/s$ total rate of gas flow is *~5 min*. The signal-delay time may be reduced to a few seconds by decreasing the diameter of the transport tubing and increasing the flow rate of the gas mixture.

The carrier gas is selected on the following basis: it is the working medium of the proportional counter and must not be activated by neutrons. Hydrogen meets these requirements, but it is flammable. Helium satisfies the latter condition. Pure helium is unfit for a counter operating in the proportional mode, and a quencher ($10\%\ CH_4$) is added to the helium at the entrance to the proportional counter for its normal operation.

To simplify the consumption control a prefabricated gas mixture of *90% He* and *10 %* $CH_4$ can be taken.

The data-acquisition system used with the counter is arranged as follows: A NIM crate with a high-voltage power supply unit and a preamplifier is placed inside the concrete shield near the counter box. A cable connects the preamplifier with the NORLAND multichannel analyzer, the CAMAC crate controller and a personal computer for data acquisition and processing, being disposed in a measuring pavilion *100 m* away. The cable resistance was matched to the input resistance of the analyzer, so that no returned signals were observed with an oscilloscope. The counter was preliminarily calibrated with a $^{109}Cd$ isotopic source by a Compton edge at an electron energy $E_e = 1.9$ *keV*, which corresponds to the gamma-ray energy $E_\gamma = 22$ *keV*. This calibration technique was selected because it is impossible to obtain a photopeak due to the gamma-rays in a helium-methane mixture. The autocalibration of the scale against a *2.9-keV* peak due to $^{37}Ar$ is used in data processing. The $^{37}Ar$ atom is an e-capture isotope; as a result a *k*-electron capture (*90%*) at an energy of *2.9-keV* is released in the proportional counter, thus producing a *k*-peak in the spectrum.

The computer code has been developed for processing of the experimental $^{37}Ar$ decay spectra and data collection. The program automatically determines the position of the maximum in the *k*-peak, which then approximates by means of a Gaussian shape and calculates the number of events under the *k*-peak with subtraction of background and $^{41}Ar$-decay events. This procedure has the property of *100%* selectivity of useful events and good background discrimination.

From Fig. 2 it is possible to evaluate the efficiency of neutron registration with the FRGM detector. The total decay speed under the $^{37}Ar$ decay plot is about $10\ sec^{-1}$. On the other hand the producing speed of $^{37}Ar$ in the detecting ampoule can be evaluated according to the relation: $I_P \sim N_{Ca}\ \sigma_{Ca}\ \Phi \sim 10^7 sec^{-1}$, where the disignations are the same as in Eq. (1) except the $N_{Ca}$ , which is the number of *Ca* atoms in the ampoule. So having in mind that at RADEX experiments the neutron flux density was $10^8$-$10^9$ $n/cm^2/sec$, the mass of powdered $CaC_2O_4$ was about *30 g* and the resulting registration efficiency was about $10^{-7}$-$10^{-6}$ which means that the FGRM detector is «transparent» to neutrons and does not disturb the neutron flux in the vicinity of the measurement area. This is a very useful property for the on-site precise neutron flux measurements. Such an efficiency makes measurements of very high neutron fluxes possible not accessible to other methods due to too high counting rates which cannot be processed by the data-acquisition system. Today's processing rate for data–acquisition is limited by the value $10^6$-$10^7$ $sec^{-1}$.

At a neutron flux of $\sim 10^9$ and a $CaC_2O_4$-mass of *30 g* the count rate of useful events in the RADEX experiment was about $10\ sec^{-1}$. Therefore, at an increase of flux up to the value of $10^{15}$-$10^{16}$ $n/cm^2/sec$, which is characterizing the neutron lifetime experiment, the mass may be lowered $10^6$-$10^7$ times at the same registration rate. This means that a point-like neutron detector with a very high neutron transparency and a very small flux disturbance can be developed. It is believed that such a kind of detector will be useful in the neutron lifetime and neuntron-neutron scattering experiments.

## 3. Nuclear physics and astrophysics

The cross-sections of the interaction of neutrons with radioactive nuclei represent the significant interest in the study of *r*- processes in stars and also the gears of isotope transmutation at irradiations in nuclear reactors. Beside the small group of rather long-lived nuclei the cross-section data on interactions of neutrons with radioactive nuclei are absent nowadays. The conventional way of obtaining such cross-sections is the time-of-flight spectrometry of neutrons. This is effective in the case of the presence of separated isotopes of radioactive nuclei. In the case of short-lived nuclei with half-life values of some days and less, obtaining of separated isotopes is technically rather complicated.

Using the flowing gas radiochemical method to register ionizing radiation can eliminate the specified difficulties. For the measurement of cross-sections of nuclear reactions with neutrons it is supposed to use the two-stage nuclear reactions, where the first stage is the formation of the investigated radioactive nucleus, and the second - the interaction of this nucleus with a neutron and formation of the radioactive isotope of an inert gas. It is obvious that in the given approach the production speed of the radioactive nuclei should be rather high for the safe registration of radioactive inert gas decays in the flowing proportional counter. The working experience obtained on the installation RADEX of INR RAS shows that an acceptable speed is close to the value of $\sim 10\ sec^{-1}$ at a background loading of $0.1\text{-}0.2\ sec^{-1}$. The evaluation of the decay speed of radioactive inert gases in the counter can be obtained by the comparison of the formation speed of the interesting radioactive nuclei $A_N$ and the radioactive isotope of the respective inert gas $A_G$:

$$A_N = N_S \sigma_1 \Phi \quad \text{and} \quad A_G = A_N t \sigma_2 \Phi\ , \qquad (2)$$

where $N_S$ is the quantity of nuclei of the initial substance in the target, $\sigma_1$, $\sigma_2$ are the cross-sections of the appropriate nuclear reactions, $\Phi$ is the flux density of neutrons, $t$ the time of irradiation of a radioactive isotope. At a weight of the target material of $\sim 100\ g$ and the cross-sections of threshold nuclear reactions with neutrons of $\sim 10\ mb$ we can obtain:

$$A_G = N_S \sigma_1 \sigma_2 t \Phi^2 \sim 10^{-27}\ t\Phi^2 . \qquad (3)$$

As can be seen from Eq. (3) it is possible to measure the cross-sections of short-lived isotopes with a half-life of $\sim 1\ sec$ and less at a neutron flux density $\Phi \sim 10^{14} n/cm^2/sec$ and higher. If the half-life of an interesting isotope is about a day and the cross-section even for the one of two nuclear reactions is about *100 mb,* the measurement of spectrum averaged cross-sections is possible at a neutron flux density of $\sim 10^{11}\ n/cm^2/sec$. This is within reach of the RADEX facility at a proton current of $1\ \mu A$. The increase of the proton current up to $10\ \mu A$ will essentially expand the spectrum of the investigated radioactive nuclei. Note that today's maximal proton current at the INR RAS linear proton accelerator is about $150\ \mu A$.

In Table 1 the data on the radioactive nuclei and on the corresponding nuclear reactions are shown, from which it is possible to extract their spectrum averaged cross-sections of the interaction with neutrons.

Table 1. The data on the radioactive nuclei and nuclear reactions.

| Rct. № | Nuclear reactions | Cross-section of radioactive nucl. formation (mb) | Half-life of radioactive nucleus | Half-life of radioactive noble gas formed | Target material |
|---|---|---|---|---|---|
| 1 | $^{10}$B(n,p)**$^{10}$Be**(n,$\alpha$n)$^{6}$He | 28 | 1.5x10$^{6}$ y | 0.8 s | |
| 2 | $^{24}$Mg(n,p)**$^{24}$Na**(n,p)$^{24}$Ne | 125 | 15 h | 3.4 m | $MgC_2O_4$ |
| 3 | $^{24}$Mg(n,p)**$^{24}$Na**(n,np)$^{23}$Ne | 125 | 15 h | 37 s | $MgC_2O_4$ |
| 4 | $^{30}$Si(n,$\alpha$)**$^{27}$Mg**(n,$\alpha$)$^{24}$Ne | 60 | 9 m | 3 m | |
| 5 | $^{30}$Si(n,$\alpha$)**$^{27}$Mg**(n,n$\alpha$)$^{23}$Ne | 60 | 9 m | 37 s | $CaC_2O_4$ |
| 6 | $^{40}$Ca(n,p)**$^{40}$K**(n,np)$^{39}$Ar | 283 | 10$^{9}$ y | 269 y | $CaC_2O_4$ |
| 7 | $^{42}$Ca(n,p)**$^{42}$K**(n,np)$^{41}$Ar | 145 | 12 h | 2 h | $CaC_2O_4$ |
| 8 | $^{43}$Ca(n,p)**$^{43}$K**(n,p)$^{43}$Ar | 100 | 22 h | 5 m | $CaC_2O_4$ |
| 9 | $^{43}$Ca(n,p)**$^{43}$K**(n,np)$^{42}$Ar | 100 | 22 h | 33 y | $CaC_2O_4$ |
| 10 | $^{44}$Ca(n,p)**$^{44}$K**(n,p)$^{44}$Ar | 20 | 22 m | 12 m | $CaC_2O_4$ |
| 11 | $^{44}$Ca(n,p)**$^{44}$K**(n,np)$^{43}$Ar | 20 | 22 m | 5 m | $CaC_2O_4$ |
| 12 | $^{46}$Ca(n,p)**$^{46}$K**(n,p)$^{46}$Ar | 12 | 105 s | 8 s | $CaC_2O_4$ |
| 13 | $^{46}$Ca(n,np)**$^{45}$K**(n,p)$^{45}$Ar | 3.3 | 17 m | 22 s | $CaC_2O_4$ |
| 14 | $^{46}$Ca(n,np)**$^{45}$K**(n,np)$^{44}$Ar | 3.3 | 17 m | 12 m | $CaC_2O_4$ |
| 15 | $^{42}$Ca(n,p)**$^{42}$K**(n,p)$^{42}$Ar | 145 | 12 h | 33 y | $CaC_2O_4$ |
| 16 | $^{48}$Ca(n,p)**$^{48}$K**(n,p)$^{48}$Ar | 14 | 7 s | ? | $CaC_2O_4$ |
| 17 | $^{48}$Ca(n,p)**$^{48}$K**(n,np)$^{47}$Ar | 14 | 7 s | 0.7 s | $CaC_2O_4$ |
| 18 | $^{48}$Ca(n,np)**$^{47}$K**(n,p)$^{47}$Ar | 0.16 | 17 s | 0.7 s | $CaC_2O_4$ |
| 19 | $^{48}$Ca(n,np)**$^{47}$K**(n,np)$^{46}$Ar | 0.16 | 7 s | 8 s | $CaC_2O_4$ |
| 20 | $^{48}$Ti(n,$\alpha$)**$^{45}$Ca**(n,$\alpha$)$^{42}$Ar | 17 | 163 d | 33 y | |
| 21 | $^{50}$Ti(n,$\alpha$)**$^{47}$Ca**(n,$\alpha$)$^{44}$Ar | 12 | 4.5 d | 12 m | |
| 22 | $^{86}$Sr (n,p)**$^{86}$Rb**(n,np)$^{85}$Kr | 22 | 18 d | 11 y | $SrC_2O_4$ |
| 23 | $^{88}$Sr (n,p)**$^{88}$Rb**(n,p)$^{88}$Kr | 6.6 | 18 m | 3 h | $SrC_2O_4$ |
| 24 | $^{88}$Sr (n,p)**$^{88}$Rb**(n,np)$^{87}$Kr | 6.6 | 18 m | 76 m | $SrC_2O_4$ |
| 25 | $^{93}$Zr(n,$\alpha$)**$^{90}$Sr**(n,$\alpha$)$^{87}$Kr | 0.5 | 29 y | 76 m | |
| 26 | $^{94}$Zr(n,$\alpha$)**$^{91}$Sr**(n,$\alpha$)$^{88}$Kr | 1.8 | 10 h | 3 h | |
| 27 | $^{93}$Zr(n,n$\alpha$)**$^{89}$Sr**(n,n$\alpha$)$^{85}$Kr | 0.5 | 50 d | 11 y | |
| 28 | $^{137}$Ba(n,p)**$^{137}$Cs**(n,p)$^{137}$Xe | 1.4 | 30 y | 4 m | $BaC_2O_4$ |
| 29 | $^{138}$Ba(n,p)**$^{138}$Cs**(n,p)$^{138}$Xe | 5 | 33 m | 14 m | $BaC_2O_4$ |
| 30 | $^{138}$Ba(n,p)**$^{138}$Cs**(n,np)$^{137}$Xe | 5 | 33 m | 4 m | $BaC_2O_4$ |
| 31 | $^{134}$Ba(n,p)**$^{134}$Cs**(n,np)$^{133}$Xe | 3.5 | 2 h | 5 d | $BaC_2O_4$ |
| 32 | $^{137}$Ba(n,np)**$^{136}$Cs**(n,np)$^{135}$Xe | 2 | 13 d | 9 h | $BaC_2O_4$ |

As in Table 1 for the case of neutrons, some two-stage nuclear reactions with protons can be proposed for the study of proton interactions with radioactive nuclei. Such kind of processes are very interesting in terms of the nucleosynthesis in the stars. This process is favored by the fact that each proton accelerator has a fixed energy, so that it is possible to measure the energy dependence of cross sections of proton interactions with radioactive nuclei. Different from this, the beams produced by neutron targets regularly have a wide spread in the neutron energy. So it is only possible to define the spectrum averaged cross sections of neutron interactions with radioactive nuclei.

Generally, in the frame of an approach proposed here, it will be very interesting to analyze the traces of the inert gases (both radioactive and stable) in the spectra of stars. Having in mind the difference in the half-life of gases, such an analysis can give useful information about nuclear reactions within the stars on the various depths, depending on the half-life of the isotopes observed. The stars have very high temperatures favoring the direct emission of inert gases from their bosom. This can give rise to a new kind of star observation – the “***noble gases astronomy***”.

It is very important to select the right structural and chemical form of the target material. This means, the target material must ensure a free emission of formed atoms of the radioactive inert gases. As a target material it is supposed to use the oxalates of the appropriate elements. Oxalates are rather heat-resistant compounds. The main advantage of oxalates is that the radioactive inert gases easily leave their powder-like structure. So far we have tested two of such chemical compounds: the oxalates of calcium and sodium. The first of the two was used for the monitoring of the neutron flux in the irradiation channel of the RADEX installation and the second for the dosimetry of secondary fast neutrons in the installation for proton therapy of cancer tumors. This is described below.

## 4. Proton therapy of cancer tumors

The purpose of activity in the given direction is the creation of a highly sensitive small-sized dosimeter for fast neutrons for the cartography of accompanying neutron fields in parts of a human body, subjected to irradiation at a proton therapy of cancer tumors. For a reliable evaluation of treatment efficiency and prevention of negative consequences of proton therapy it is necessary to know a radioactive dose obtained by various patient organs as a results of effects of all kinds of ionizing radiation. The devices for the individual formation of the therapeutic beam located in immediate proximity from an irradiated organ absorb a sizeable part of dropping protons, in this way producing secondary neutrons. The flux of these neutrons falling on the vital organs of a patient can have dangerous consequences for the health of a patient. Obviously, the operative control of human organs’ irradiation with secondary neutrons can be executed only by the use of compact and rather fast-response detectors of fast neutrons. Therefore the development of new methods of neutron registration remains the actual problem.

For high safety at the proton therapy of tumors it is very important to receive the information on the distribution of fluxes of accompanying fast secondary neutrons in organs of a human body subjected to irradiation in a real time mode. Consequently it is necessary always to have the best possible spatial resolution. The given circumstances force to minimize the size of a detecting element of the neutron dosimeter, which in an ideal case should have dot size. Except the secondary neutrons at the proton irradiation at an energy of *~200 MeV*, a great amount of gamma-quanta arises, which can essentially worsen the background conditions. As the biological efficiency of irradiation by neutrons and other particles can essentially differ from each other, it is desirable to have a specific neutron detector. Besides it is important to protect the electron acquisition system of the detector against the effect of all kinds of radiation.

The help of the FGRM detector can largely solve the specified problems. For a case history of the method we shall consider all possible nuclear reactions of neutrons producing the radioactive isotopes of inert gases. The list of the appropriate nuclear reactions is adduced in Table 2.

Table 2. Characteristics of neutron nuclear reactions yielding radioactive isotopes of inert (noble) gases.

| Rct. № | Nuclear reaction | Nat. isotopic abundance ( % ) | Half-life of gaseous product | Nuclear reaction threshold (MeV) | Cross-section at 14.5 MeV (mb) |
|---|---|---|---|---|---|
| 1 | $^{6}Li(n,p)^{6}He$ | 7.5 | 0.8 s | 3.2 | 5.8 |
| 2 | $^{7}Li(n,np)^{6}He$ | 92.5 | 0.8 s | 11.4 | 10 |
| 3 | $^{7}Li(n,d)^{6}He$ | 92.5 | 0.8 s | 9.0 | 11 |
| 4 | $^{9}Be(n,\alpha)^{6}He$ | 100 | 0.8 s | 0.7 | 10 |
| 5 | $^{23}Na(n,p)^{23}Ne$ | 100 | 37 s | 3.8 | 60 |
| 6 | $^{26}Mg(n,\alpha)^{23}Ne$ | 11.1 | 37 s | 7.0 | 85 |
| 7 | $^{39}K(n,p)^{39}Ar$ | 93.3 | 269 y | 1.0 | 385 |
| 8 | $^{41}K(n,p)^{41}Ar$ | 6.73 | 1.83 h | 3.0 | 8.5 |
| 9 | $^{40}Ca(n,\alpha)^{37}Ar$ | 96.94 | 35 d | 1.0 | 130 |
| 10 | $^{42}Ca(n,\alpha)^{39}Ar$ | 0.6465 | 269 y | 2.5 | 50 |
| 11 | $^{44}Ca(n,\alpha)^{41}Ar$ | 2.09 | 1.83 h | 6.0 | 36 |
| 12 | $^{46}Ca(n,\alpha)^{43}Ar$ | 0.0035 | 5.37 m | 6.0 | 5 |
| 13 | $^{48}Ca(n,\alpha)^{45}Ar$ | 0.19 | 21.5 s | 14.0 | 1.6 |
| 14 | $^{85}Rb(n,p)^{85}Kr$ | 72.17 | 10 y | 2.5 | 18 |
| 15 | $^{87}Rb(n,p)^{87}Kr$ | 27.83 | 76 m | 3.1 | 10 |
| 16 | $^{88}Sr(n,\alpha)^{85}Kr$ | 82.6 | 10 y | 9.0 | 6 |
| 17 | $^{133}Cs(n,p)^{133}Xe$ | 100 | 5.25 d | 6.0 | 11 |
| 18 | $^{138}Ba(n,\alpha)^{135}Xe$ | 71.7 | 9 h | 7.5 | 3 |

As can be seen from Table 2 the half-life values of the produced radioactive inert gases differ essentially from each other. To evaluate the sensitivity of a selected nuclear reaction from Table 2 against a neutron flux it is possible to take advantage of the Eq. (1). Fig. 5 represents the relation of the decay rate of the radioactive inert gas in the counter versus the fast neutron flux on an ampoule. Each shaded area corresponds to one of the nuclear reactions from Table 2. The spread of sensitivity within the size of the shaded area is stipulated by the change of the weight of target material in an ampoule within the range of *1-10 g*. All remaining parameters included in Eq. (1) are fixed and identical to all nuclear reactions. The upper limit of the decay speed is about $10^{6}\ sec^{-1}$, which is near the ultimate capability of electronic data acquisition. As it is visible from Fig. 5, the greatest sensitivity has the *reaction 4,* whereas the sensitivity of *reaction 5* is slightly lower. It is believed that it is possible to create the sensitive dosimeter for fast neutrons using these two nuclear reactions. However, the specified reactions cannot be used in the case of high-density neutron fluxes (for example, in an active zone of nuclear reactors or neutron targets), since the counting speed will be too high (electronics engineering " will choke "). For very high-density neutron fluxes the reactions with smaller sensitivity will approach, for example, the three *reactions 13*, *9* and *7* of Table 2.

The short half-life of $^6He$, in accordance with Eq. (1), allows increasing the sensitivity for small quantities of target material in the detecting ampoule substantially. The counting at a constant time of gas exchange of $\tau \sim 1\ sec$ can be compared to the formation speed of $^6He$ in the nuclear *reaction 4* of Table 2:

$$A = N_{Be}\sigma_{Be}\Phi\ . \qquad (4)$$

At a weight of $Be \sim 10\ g$ the total number of atoms in the ampoule is $N_{Be} = 6x10^{23}$. Having in mind $\sigma_{Be} = 100\ mB$, for the count speed we can obtain $A \sim 6$ *decays/sec* at the neutron flux density $\Phi \sim 100\ n/cm^2/sec$. At $\tau \sim 1\ sec$, $V_C \sim 10\ cm^3$, $V_a \sim 1\ cm^3$ we will obtain a consumption of transport gas of $L \sim 10\ cm^3/sec$, which satisfies the condition of applicability of Eq. (1): $Va/L << \tau$. Such a decay speed is acceptable, because the background count in the vicinity of the proton therapy installation was during the test measurements $\sim 0.1\ sec^{-1}$. Beside there is the possibility to increase the measurement time of the neutron flux in one point up to, for example, *10 sec*. Thus the count speed will be $60\ sec^{-1}$, which is *5* times faster than we had in the experiment with a *reaction 9* of Table 2 at the installation RADEX.

It is also possible to realize the dosimeter on the base of *reaction 5* of Table 2: $^{23}Na\ (n, p)\ ^{23}Ne$. Though the half-life of $^{23}Ne$ is *~40* times higher than those for $^6He$, nevertheless, the sensitivity of the dosimeter will change insignificantly, since the cross-section of a *reaction 5* is *6* times higher than the *reaction 4*, besides the sensitivity can be increased at the expense of decreasing the consumption of transport gas as it follows from the Eq. (1). This leads however to an increase of the gas exchange constant, but it is permissible due to the greater half-life of $^{23}Ne$. The advantage of a *reaction 5* as compared with the *reaction 4* may turn out to be the more convenient way of decay registration of a radioactive inert gas, because apart from the beta-decay $^{23}Ne$ produces gamma-quanta, which can be registered by a usual semiconductor.

## 5. Nondestructive control of fission materials

The development of new methods of neutron registration remains the actual problem for the control of radioactive materials movement. In case of large freight traffic and conveyer system of freight examination at the transfer points, customs terminals and junctions it is important to create a compact and transferable dosimeter which is capable of conducting the nondestructive express-analysis of the presence of radioactive materials in cargo containers and luggage of probable terrorists at a small exposure time in the real-time scale. Thus it is necessary to have, whenever possible, the greatest sensitivity to register minimum quantities of radioactive substances, especially of fissionable materials of various chemical forms. To solve the problem of developing an express monitor for the nondestructive detection of $\alpha$-emitting isotopes of trans-uranium elements it is proposed to measure neutrons from $(\alpha,n)$-reactions on the light elements using a FGRM detector. The neutrons formed in the given reactions are supposed to be registered with the help of the *reactions 4* and *5* of Table 2 using the target material of beryllium- or sodium-oxalate correspondigly.

For the nondestructive control of fission materials it is important to minimize the mass of target material. Then it is also possible to use the other not so sensitive nuclear reactions of Table 2. One can imagine an installation of the nondestructive control in the airports as follows: the conveyer passes the luggage through a frame-like tube filling with powder of target material. The length of the frame perimeter would be *~3 m*. At a tube diameter of *~2 cm* the total mass of the target substance in the frame would be *~500 g*.

To evaluate the detector sensitivity we shall assume that the source of neutron radiation represents a homogeneous mixture of an alpha-emitting trans-uranium isotope with any light element, i.e. *O, F, C, Al,* etc. The yield of fast neutrons from such a composition can be set equal to $10^{-6}$ per alpha particle for an evaluation. For the sensitivity of the device it is then possible to obtain the following values: at a distance of *0.5 m* from the container, the device is capable to register *40 g Th-230, 20 g Pu-239, 2 g Pu-240, 0.2 g Am-241 or 0.004 g Cf-250* during an exposition time of *1* minute. By increasing the mass of target material the detector sensitivity increases linearly.

## 6. Monitoring of fluxes and measurement of spectra of bombarding particles

For the monitoring of fast neutron fluxes at the installation RADEX the nuclear *reaction 9* (on $CaC_2O_4$) from Table 2 was used. It has been shown earlier [14] that $^{37}Ar$ leaves the crystalline lattice of the water-free calcium oxalate at zero activation energy.

With the help of FGRM the neutron spectrometry using several nuclear reactions from Table 2 can be realized with various energy thresholds. It is interesting to note that on calcium of a natural composition the four nuclear *reactions (9 - 13)* are possible with the various thresholds. This means that the simple spectrometry of fast neutrons can be realized on the one target material: the oxalate of calcium of natural composition.

Apart from the neutrons the FGRM allows to register elementary particles and also protons, deuterons, tritons and heavy nuclei. In Table 3, as an example, some nuclear reactions with protons are collected producing radioactive isotopes of inert gases.

Table 3. Proton-induced nuclear reactions producing radioactive inert (noble) gases.

| Rct. № | Nuclear reaction | Nat. isotopic abundance ( % ) | Half-life of gaseous product | Energy threshold (MeV) | Maximal cross-section (mb) |
|---|---|---|---|---|---|
| 1 | $^{19}F(p,n)^{19}Ne$ | 100 | 17.2 s | 4.0 | 70 |
| 2 | $^{37}Cl(p,n)^{37}Ar$ | 24.23 | 35 d | 1.5 | 300 |
| 3 | $^{79}Br(p,n)^{79}Kr$ | 50.7 | 35 h | 4.0 | 900 |
| 4 | $^{79}Br(p,3n)^{77}Kr$ | 50.7 | 74.7 m | 15 | 250 |
| 5 | $^{79}Br(p,4n)^{76}Kr$ | 50.7 | 14.8 h | 35 | 45 |
| 6 | $^{127}I(p,n)^{127}Xe$ | 100 | 36.4 d | 3.0 | 500 |
| 7 | $^{127}I(p,3n)^{125}Xe$ | 100 | 17.0 h | 20 | 850 |
| 8 | $^{127}I(p,5n)^{123}Xe$ | 100 | 2.08 h | 40 | 500 |
| 9 | $^{127}I(p,6n)^{122}Xe$ | 100 | 20.1 h | 50 | 125 |
| 10 | $^{127}I(p,7n)^{121}Xe$ | 100 | 40.1 m | 55 | 90 |

From Table 3 the principal capability of proton spectrometry in the energy range of 1.5-55 *MeV* can be seen. At present two target materials are known which can be used for proton monitoring, namely, $C_6F_6$ and $C_6Cl_6$ for the nuclear *reactions 1* and *2* of Table 3, correspondingly. Both of the two are powder-like substances allowing the free emission of the corresponding radioactive inert gases.

## 7. Diagnostics of thermonuclear plasma

From the point of view of the capabilities of a FGRM explained above the monitoring of neutrons at an energy of 14 *MeV* in the thermonuclear installations is possible, practically by use of any of the nuclear reactions listed in Table 2. However, the given method cannot be used when the monitor should give a control action during the time smaller than the minimum delay time of a signal in a FGRM. At present it is possible to consider the value of a few seconds as a minimum delay time of a signal.

From a point of view of thermonuclear plasma the *reactions 12* and *13* of Table 2 are potentially interesting. In principle it is possible to optimize an isotope composition of the target, so that the nuclear reactions $^{46}Ca\ (n,\alpha)^{43}Ar$; $^{48}Ca\ (n,\alpha)^{45}Ar$ with the energy thresholds of *6* and *14 MeV,* correspondingly, can be used for temperature diagnostics of thermonuclear plasma, for example, in (*d-t)*-experiments DTE1 [15]. Actually the cross-section of the *reaction 13* gains the non-zero value exactly at the energy of *14 MeV* and further grows linearly with the increase of the energy (see Fig.6). As shown in the reference [16] in the case of Maxwell's plasma, the neutron spectrum in the vicinity of *14 MeV* has the Gaussian shape. Thus the width of energy distribution of a neutron pulse $\Delta E_n$ is connected with the plasma temperature $T_p$ as follows:

$$\Delta E_n = 178\sqrt{T_P}\ , \qquad (6)$$

where the values of $\Delta E_n$ and $Tp$ are expressed in *keV*. Thus, with the help of *reaction 13* we can determine a half-width of Gaussian distribution of a neutron pulse of a (*d-t)*-reaction, and, hence, the temperature of plasma from Eq. (6). To be sure that the given widening of a neutron pulse is connected just to the plasma temperature the *reaction 12* of Table 2 can be used: this cross-section is constant in the energy range of *8-20 MeV*. For a realization of the specified way of measurement of the plasma temperature it is necessary to have two detectors: one of these uses the partition isotope *Ca-48* as a target material, and the second *Ca-46*. Thus the contents of other isotopes of calcium should be lowered up to a minimum, so that the competing reactions could not affect the results of measurement.

## 8. Conclusions

In this work some applications of a flowing gaseous radiochemical method (FGRM) for the registration of ionizing radiation are considered. The "brain storm" carried out by the authors of this work and active discussions of the given method with the help of Russian and international experts [2-4] have allowed to outline a rather broad field of probable applications including:

1. Monitoring of neutron flux time behavior in the neutron lifetime experiment.
2. Measurement of cross-sections of the interaction of neutrons and protons with radioactive nuclei, important for nuclear physics and astrophysics.

3. The creation of the compact positional-sensing dosimeters of fast neutrons necessary for the therapy of cancer tumors at the installations driven by linear proton accelerators.
4. Remote, nondestructive control of fissionable materials at the airports and freight examination on transfer points, customs terminals and junctions.
5. Monitoring of fluxes and measurement of spectra of neutrons, protons, other elementary particles, and also light and heavy ions and radioactive nuclei.
6. Measurement of the plasma temperature in the thermonuclear installations.

## <u>List of the 3 tables</u> *(integrated in the text)*

## <u>List of the 6 Figures</u>

Fig. 1. Cross sectional view of the RADEX facility.

Fig. 2. The decay spectrum in the proportional counter during the experiments with the RADEX target.

Fig. 3. The time history of a neutron flux monitoring at the RADEX target with the help of the FGRM neutron detector. As target material $CaC_2O_4$ was used. The dark colored areas correspond to the presence of a proton beam on the target.

Fig. 4. The scheme of the gaseous tract in the experiment on the neutron flux monitoring at the RADEX target.

Fig. 5. The relation of the decay speed of radioactive inert (noble) gases as a function of a fast neutron flux on the detecting ampoule. Each of the shaded areas is marked with the number corresponding to one of the neutron reactions of Table 2. The spread of sensitivity within the width of a shaded area corresponds to the variation of the mass of target substance within the range of 1-10 g.

Fig. 6. Cross sections of formation of $^{46}Ca$ and $^{48}Ca$ as a function of neutron energy.

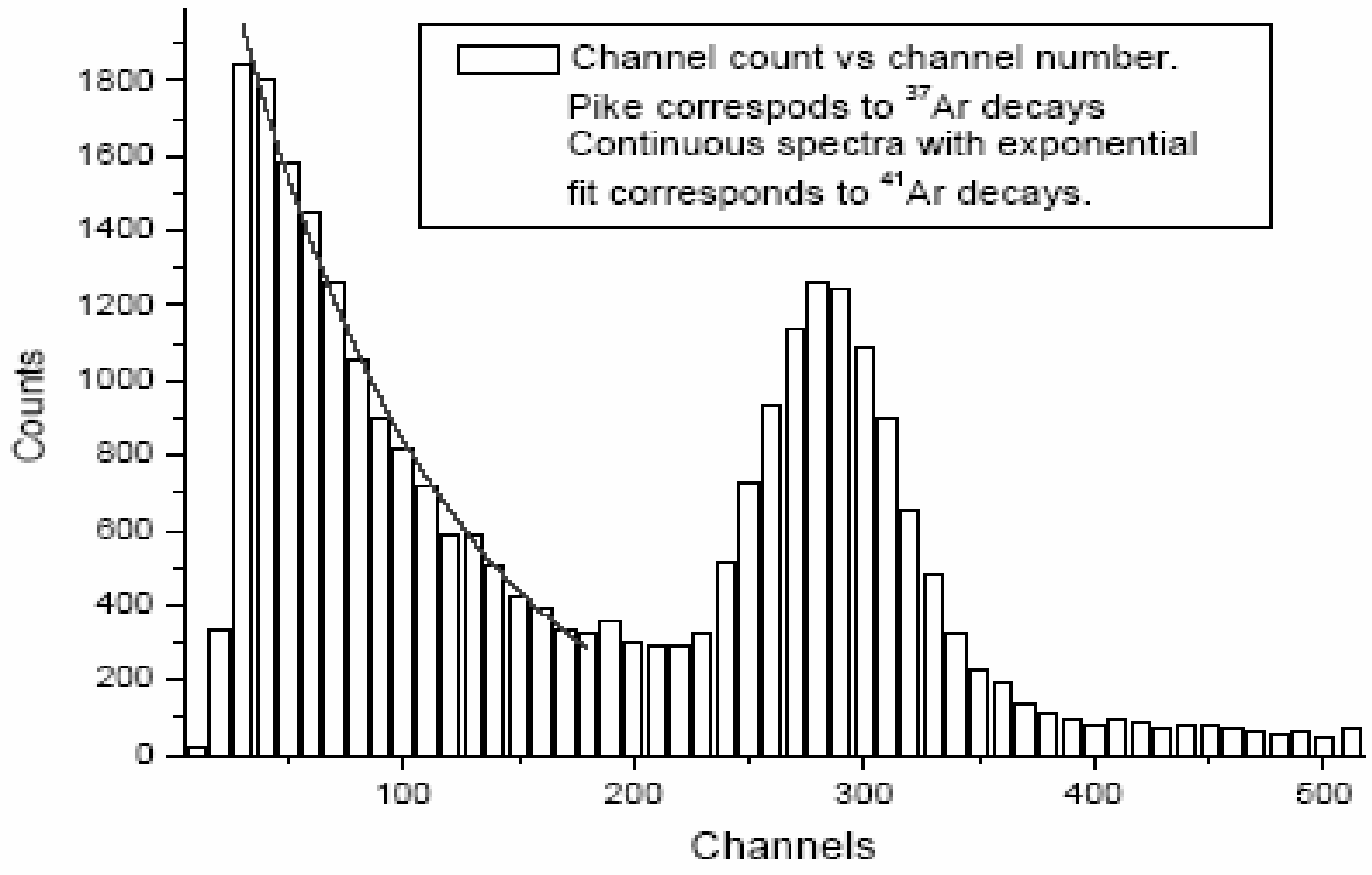

Fig.2. The decay spectrum in the proportional counter during the experiments with RADEX target.

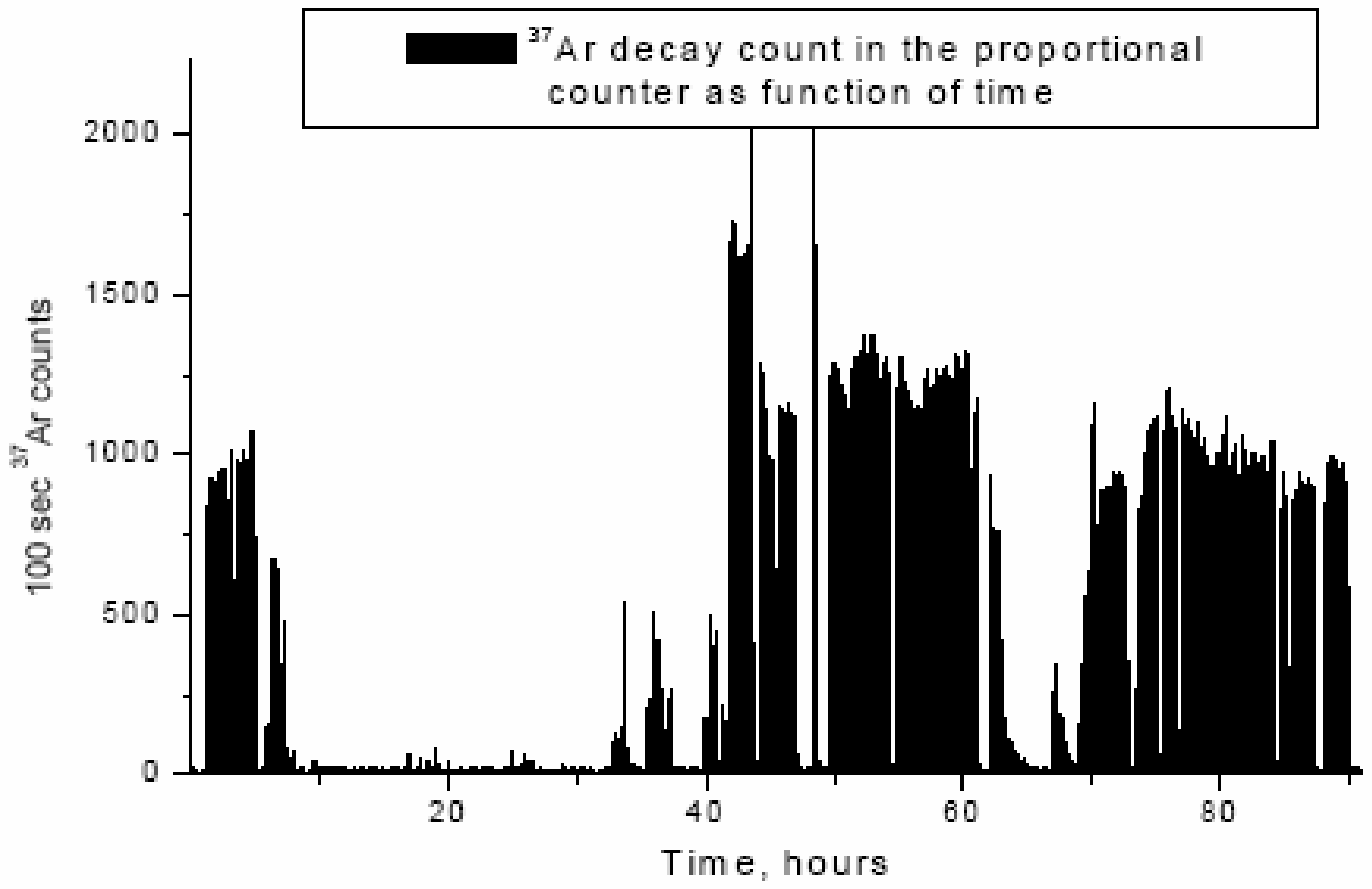

Fig.3. The time history of a neutron flux monitoring at the RADEX target with the help of FGRM neutron detector. As the active substance the $CaC_2O_4$ was used. The dark color areas corresponds to the presence of a proton beam on the target.

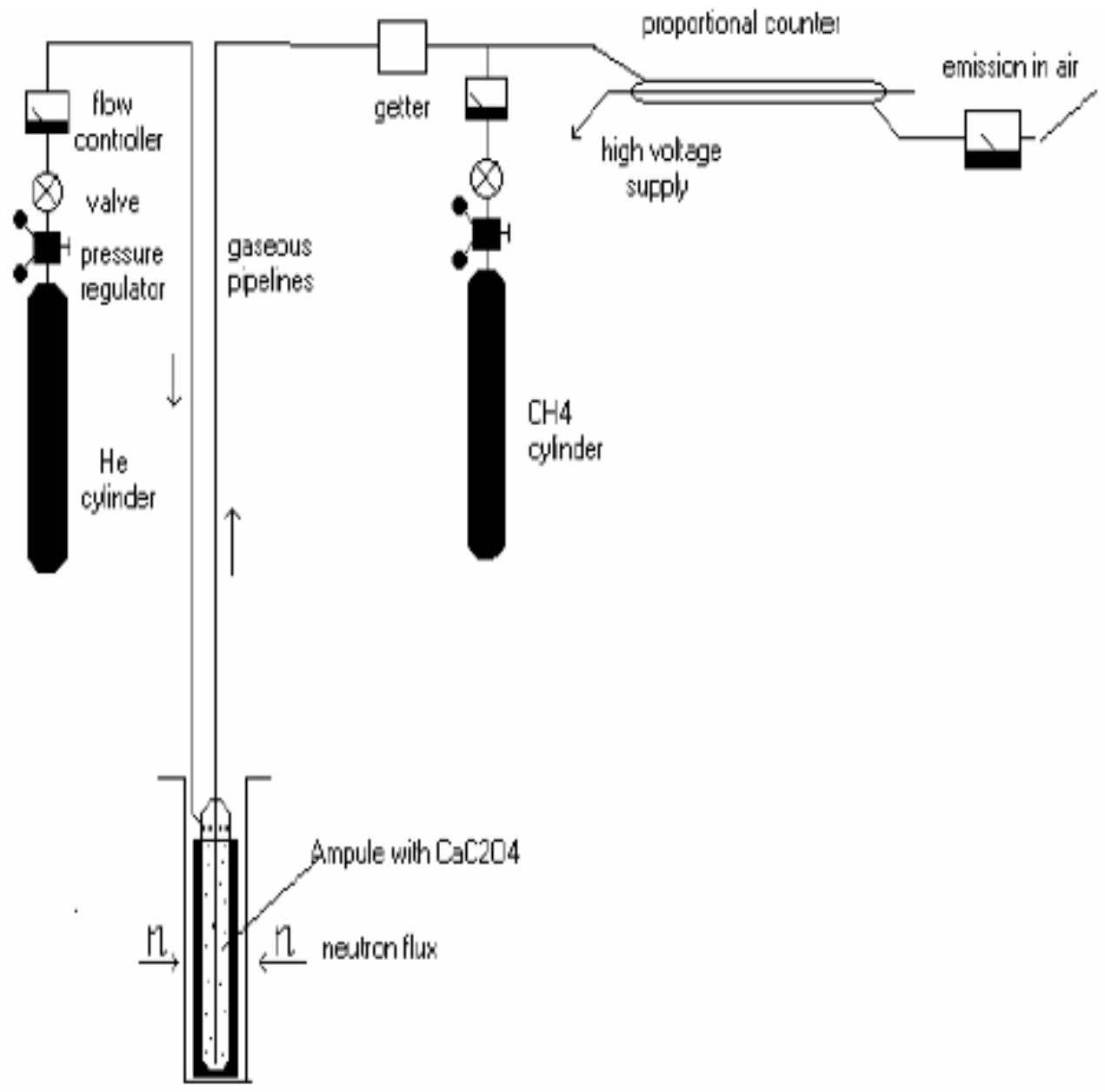

Fig.4. The scheme of the gaseous tract in the experiment on neutron flux monitoring at the RADEX facility.

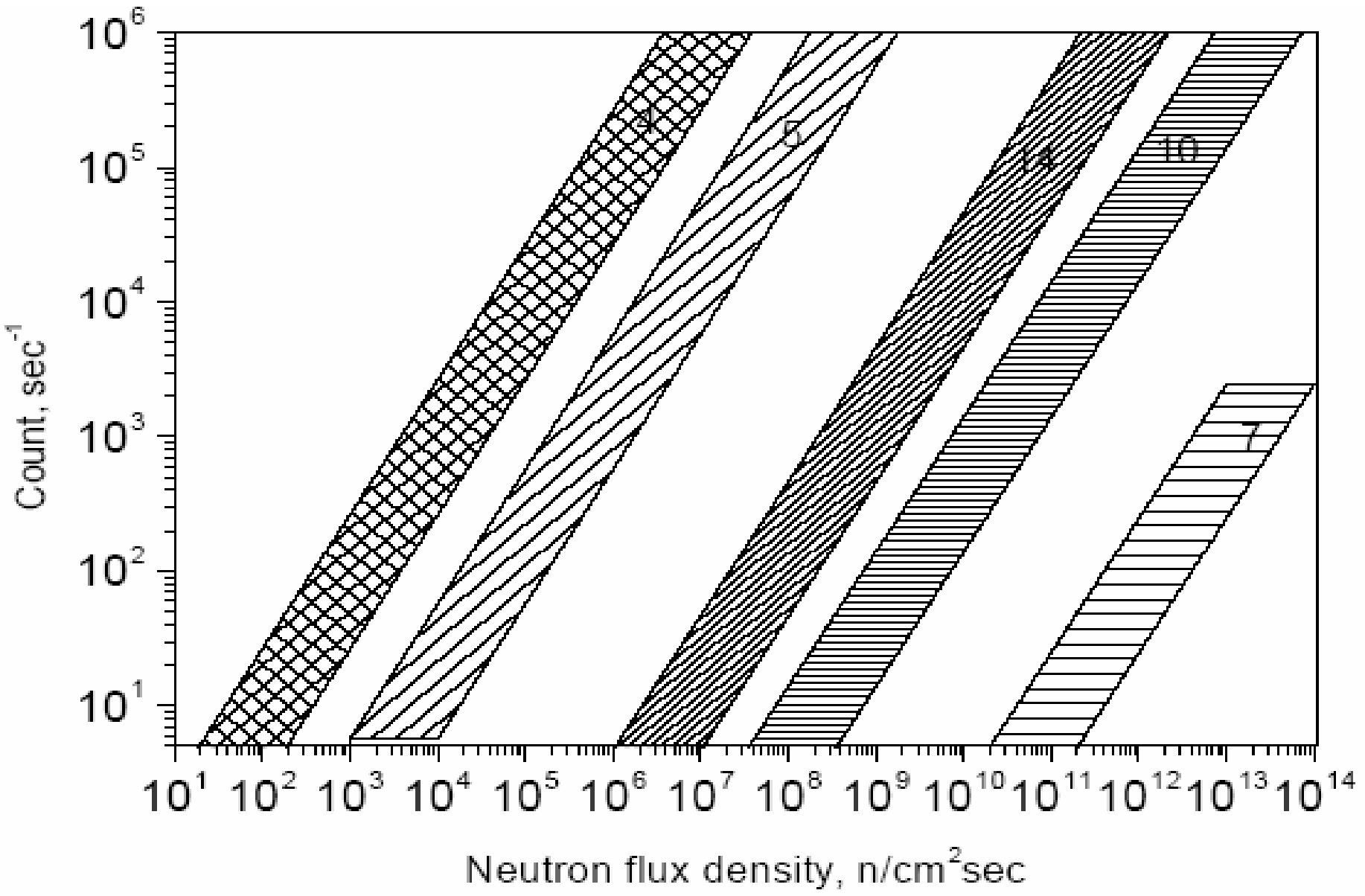

Fig.5. The relation of decay speed of noble radioactive gases as a functions of fast neutron flux on the detecting ampoule. Each from the shaded areas is marked with the figures corresponding to one of the nuclear reaction of a Table 2. The spread of sensitivity within the width of shaded area corresponds to variation of the mass of active substance within the range of 1- 10g.

**Fig.6. The cross sections of formation of Ca -46 and Ca-48 as a function of neutron energy**

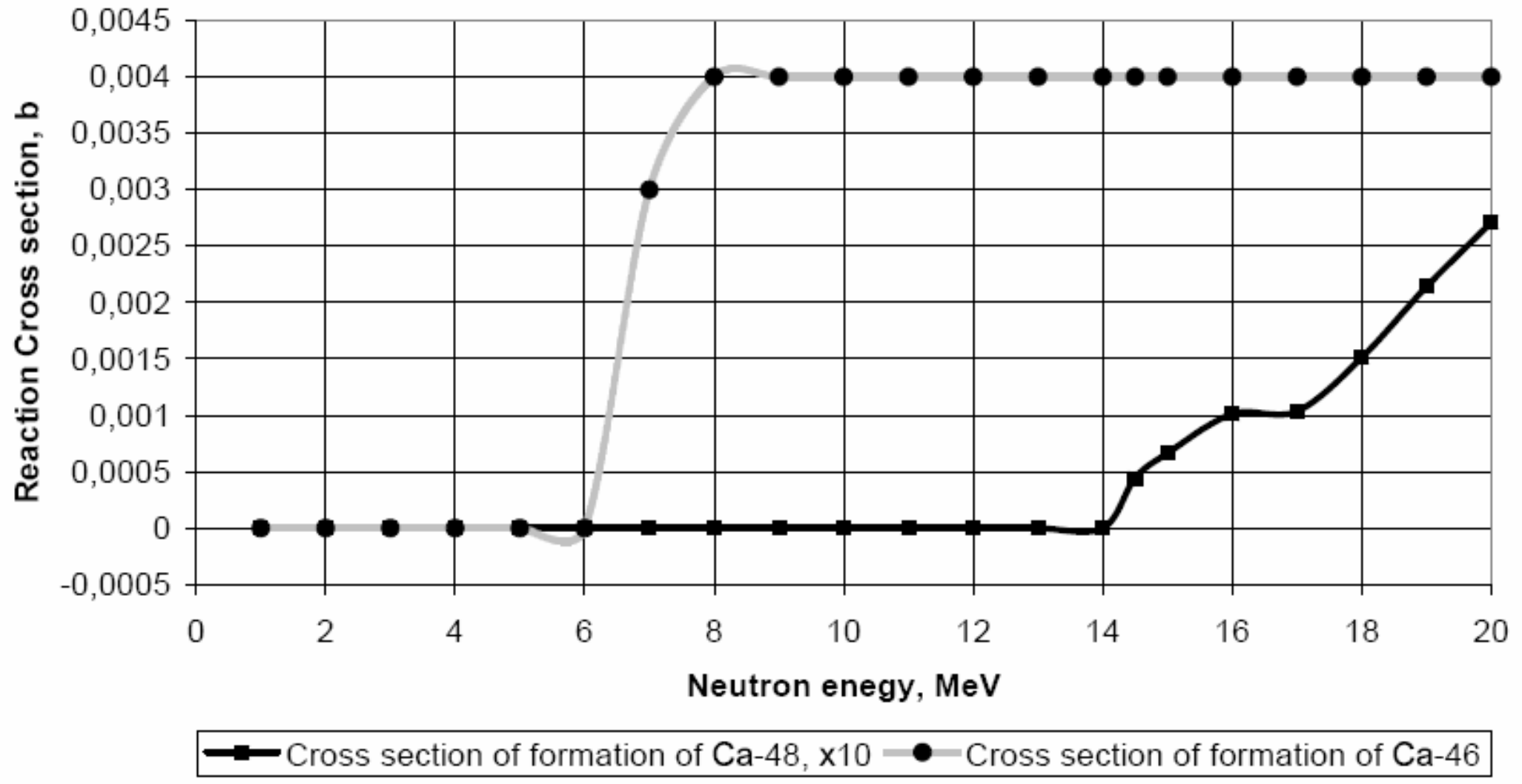